\def\be{\begin{equation}}
\def\ee{\end{equation}}
\def\bea{\begin{eqnarray}}
\def\eea{\end{eqnarray}}

\def\half{\frac{1}{2}}

\def\ij{{\langle ij \rangle}}
\def\bj{{ \bf j }}
\def\bk{{\bf  k }}

\def\OM{\Omega}
\documentstyle[12pt]{article}
\begin{document}
\title{Spin Tunneling, Berry phases
and Doped Antiferromagnets}

\author{Assa Auerbach\\
Department of Physics,\\
 Technion, Haifa 32000, Israel.}
\maketitle


Interference effects between Berry phase factors in spin tunneling systems have
been
discussed in recent Letters by Loss, DiVincenzo and Grinstein (LDG) \cite{LDG}
and von Delft and Henley (vDH) \cite{vDH}. As physical realizations, LDG
proposed  tunneling of
magnetic clusters, and vDH proposed an anisotropic spin hamiltonian  and the
Heisenberg antiferromagnet on the Kagom\'e lattice. In these examples, for
certain spin sizes $s$, the classical ground state degeneracies are not lifted
by tunneling processes.

This Comment points out that Berry phases in spin tunneling are important in
another interesting case: the two dimensional doped antiferromagnet. This model
is often used to describe cuprate superconductors.
We shall see that the
Berry phases of Refs. \cite{LDG,vDH} can explain the ground state momentum of
the single hole, as found by finite size numerical simulations of the t-J model
\cite{D}.

In Ref. \cite{AL}, the hole in the antiferromagnet has been cast as a spin
tunneling problem. First the t-J model was extended to $s > \half$, which was
then described in a spin coherent states path integral by the   hamiltonian
function
\be
H~= J\sum_\ij \OM_i\cdot \OM_j (1-\rho_i)(1-\rho_j)~ - E^{hole}  [\OM;t]
\label{1}
\ee
$t$ is the hole hopping parameter.
$\rho_i$ and $E^{hole}$ are the hole density and energy.
In the large $s$ limit, the path integral can be expanded about the ground
states of $H$, and $E^{hole}$ and $\rho$ can be approximted by adiabatic
functions of the instanateous spin configurations.

The semiclassical expansion yields the following results \cite{AL}: For $1<t/J
<4.1$,  (\ref{1}) is minimized by the five site polaron
(one flipped spin at site $a$).   Translational invariance  is restored by {\em
spin tunneling  paths}, and the polaron acquires a disperion relation
\be
\epsilon_\bk~= \sum_j e^{i\bk \cdot \bj} \Gamma_{ij}
\label{2}
\ee
where the tunneling rates $\Gamma_{ij}$ vanish for $i,j$ on different
sublattices.
As for a particle in a double well, in the absence of Berry phases
$\Gamma_{ij}$ is negative. Here we shall see that {\em for half odd integer
spins} one obtains an extra minus sign, such that $\Gamma >0$, thereby shifting
the ground state momentum from $\bk=(0,0),(\pi,\pi)$ to the line $\cos k_x +
\cos k_y =0$.

The sign of $\Gamma_{ab}$ is determined as follows. Consider a polaron in a
local antiferromagnetic order in the ${\hat x}$ direction. The tunneling path
is parametrized by $\varphi$: the average angle of rotation of spins $a$ and
$b$. To satisfy energy conservation, $H[\OM]-E=0$, we let the $z$ components be
imaginary, i.e. $\cos\theta_i~\to -i{\tilde\mu}_i$. Following
Ref. \cite{vDH} (see also Refs. \cite{GK,AK}), and summing over the clockwise
and anticlockwise rotations of $\varphi$ we obtain
\be
\Gamma_{ab} =-  {\cal N}
\exp\left(-\int_0^\pi d\varphi
\sum_i (s-{\rho_i\over 2}) \mu_i{d\phi_i\over d\varphi}\right)
\sum_\pm e^{i\Upsilon^{\pm} }
\label{4.0}
\ee
${\cal N}$ is the fluctuation prefactor, and the Berry phases are
\be
\Upsilon^{\pm} = \int_0^{\pm\pi} d\varphi
\sum_i (s-{\rho_i\over 2}) {d\phi_i\over d\varphi}
\label{4}
\ee
The term of order $s$ is simply $\pm 2\pi s$, due to the rotations of $\phi_a$
and $\phi_b$. Other spins retrace their paths and their contribution to that
term vanishes.  Thus,
\be
-\sum_\pm e^{i\Upsilon^{\pm} }=- e^{i 2\pi s} \cos\left[\sum_i \int_0^{\pi}
d\varphi {\rho_i\over 2} {d\phi_i\over d\varphi}\right]
\ee
which yields
\be
\Gamma~=- e^{i2\pi s} |\Gamma|
\label{5}
\ee
Therefore the dispersion of a single hole in the half odd integer
antiferromagnet is {\em  maximized} at  $\bk=(0,0),(\pi,\pi)$ in agreement with
Ref. \cite{D}. For integer $s$ however,
Eqs. (\ref{2},\ref{5})  predict the ground state of the t-J model to be at
$\bk=(0,0)$ or $\bk=(\pi,\pi)$.

I thank C. Henley and A. Garg for helpful conversations. The hospitality of the
Aspen Center for Physics is gratefully acknowledged.  This work is supported by
the U.S-Israeli Binational Science Foundation
Grant No. 90-00441-1 and the Fund for the Promotion of Research at the
Technion.

\end{document}